\documentclass[twocolumn,prc,amsmath,amssymb,nofootinbib,preprintnumbers]{revtex4}

\usepackage{color}
\usepackage{graphicx}

\usepackage{dcolumn}
\usepackage{bm}
\usepackage{lscape}
\usepackage[normalem]{ulem}

\usepackage{lipsum}

\usepackage{txfonts}
\usepackage{mathrsfs}

\begin{document}
\preprint{NITEP 200}
\title{Investigation of the determination of nuclear deformation using
 high-energy heavy-ion scattering}

\author{Shin Watanabe}%
\email{s-watanabe@gifu-nct.ac.jp}
\thanks{This work was partly conducted during a sabbatical leave at the Departamento de 
F\'isica At\'omica, Molecular y Nuclear, Universidad de Sevilla.}
\affiliation{National Institute of Technology (KOSEN), Gifu College, Motosu 501-0495, Japan}
\affiliation{RIKEN Nishina Center, Wako 351-0198, Japan}

\author{Takenori Furumoto}
\email{furumoto-takenori-py@ynu.ac.jp}
\affiliation{College of Education, Yokohama National University, Yokohama 240-8501, Japan}

\author{Wataru Horiuchi}%
\affiliation{Department of Physics, Osaka Metropolitan University, Osaka 558-8585, Japan}
\affiliation{Nambu Yoichiro Institute of Theoretical and Experimental Physics (NITEP), Osaka Metropolitan University, Osaka 558-8585, Japan}
\affiliation{RIKEN Nishina Center, Wako 351-0198, Japan}
\affiliation{Department of Physics, Hokkaido University, Sapporo 060-0810, Japan}

\author{Tadahiro Suhara}%
\affiliation{National Institute of Technology (KOSEN), Matsue College, Matsue 690-8518, Japan}

\author{Yasutaka Taniguchi}%
\affiliation{Department of Computer Science, Fukuyama University, Fukuyama 729-0292, Japan}
\affiliation{RIKEN Nishina Center, Wako 351-0198, Japan}
\affiliation{National Institute of Technology (KOSEN), Kagawa College, Mitoyo 769-1192, Japan}

\date{\today}

\begin{abstract}
\noindent{\bf Background:}
Nuclear deformation provides a crucial characteristic of nuclear structure.
Conventionally, the quadrupole deformation length 
of a nucleus, $\delta_{2}$, has often been determined  based on a macroscopic model through a deformed nuclear potential with the  
deformation length $\delta^{\rm (pot)}_{2}$, which
 is determined to reproduce the nuclear scattering data.
This approach assumes $\delta_{2}=\delta^{\rm (pot)}_{2}$ although there is no theoretical foundation.

\noindent{\bf Purpose:}
We clarify the relationship between $\delta_{2}$ and 
$\delta^{\rm (pot)}_{2}$ for high-energy heavy-ion scattering systematically to evaluate the validity of the conventional approach to determine the nuclear deformation.

\noindent{\bf Method:}
The deformation lengths for the $^{12}$C inelastic scattering by $^{12}$C, $^{16}$O, $^{40}$Ca, and $^{208}$Pb targets at $E/A$ = 50--400 MeV are examined.
First, we perform microscopic coupled-channel (CC) calculations to 
relate $\delta_{2}$ of the deformed density 
into the inelastic scattering cross section.
Second, we use the deformed potential model to determine $\delta^{\rm (pot)}_{2}$ so as to reproduce the microscopic CC result.
We then compare $\delta^{\rm (pot)}_{2}$ with $\delta_{2}$.

\noindent{\bf Results:}
We find that $\delta^{\rm (pot)}_{2}$ is about 20--40 \% smaller than presumed $\delta_{2}$, showing strong energy and target dependence.
Further analysis, which considers higher-order deformation effects beyond
the derivative model, reveals that $\delta^{\rm (pot)}_{2}$ is still about 15--35 \% smaller than $\delta_{2}$.

\noindent{\bf Conclusion:}
Our results suggest that one needs to be careful
when the deformed potential model for the high-energy heavy-ion scattering is used to extract the nuclear deformation. 
The conventional approach may underestimate the deformation length $\delta_2$ systematically.
\end{abstract}

\maketitle
\section{Introduction}
The determination of nuclear deformation is one of the key issues in nuclear physics. As it significantly influences the nuclear structure and reaction dynamics~\cite{BM,DNR}, the nuclear deformation has extensively been investigated from various points of views~\cite{MAT83, HAO84, KHO00, IWA00, TAK09, LEE09, FAL13, MIC14, CAN15, DOO16, KUN19, JIA20, HOL21, CHE22, KUN22}.
The nuclear surface of an axially symmetric deformed nucleus is often represented as
$R(\theta' )= R_0 \left[ 1 + \sum_\lambda \beta_{\lambda} Y_{\lambda 0}(\theta' ) \right]$, 
where $R_0$ and $\beta_{\lambda}$ denote the radius parameter and the deformation parameter with multipolarity $\lambda$, respectively.
Determining the deformation length $\delta_\lambda = R_0 \beta_{\lambda}$ is important as it
provides a crucial indicator of nuclear deformation. 
The deformation length is also an essential input to the collective model, which offers a simple and powerful description
of atomic nuclei, allowing us to predict the electromagnetic properties as well as the inelastic scattering cross sections
with the help of the distorted wave Born approximation (DWBA).

In the present work, we focus on the most basic nuclear deformation, quadrupole deformation.
The quadrupole deformation length $\delta_2$ 
of a nucleus has often been deduced from inelastic scattering cross sections using
the conventional approach such as coupled-channel (CC) formalism and DWBA.
These conventional approaches are based on the collective model,
and are often referred to as the deformed potential (DP) model~\cite{KHO00}.
In the DP model, the deformation length $\delta^{\rm (pot)}_{2}$
of the nuclear optical potential is determined to reproduce
the experimental cross sections.
In order to extract $\delta_{2}$ in the DP model, the relation
\begin{equation}
\delta_{2}=\delta^{\rm (pot)}_{2}\label{eq:delta2}
\end{equation}
is often assumed.
However, this assumption has no basis
and is questionable because $\delta^{\rm (pot)}_{2}$
includes the information on both the projectile and target nuclei, and also the nuclear force.
Based on this unestablished assumption~\eqref{eq:delta2}, 
$\delta_{2}$ has been experimentally determined
with the DP model~\cite{DOO14, LI15, VAQ19, WIM21, REV23}.

Here, we take a microscopic approach for extracting $\delta_{2}$.
Over the past five decades, a microscopic CC calculation for heavy-ion scattering has been developed significantly~\cite{SAT79, DNR} and
has been widely used to investigate nuclear structure and reactions~\cite{DNR, SAK86, SAT97, KHO00, KAT02, KHO02, KHO05, FUR13-2, FUR18, FUR21-1, FUR21-2,MIN16}.
These calculations are based on the double-folding model, where
the nuclear optical potential is constructed by folding the effective nucleon-nucleon interaction with the projectile and target densities.
When a coupling potential is required, a transition density, which reflects the deformation effect, is incorporated into the folding procedure.
Henceforward, we call this microscopic framework the deformed density (DD) model
to distinguish it from the DP model.
The DD model enables us to extract $\delta_{2}$ directly, not via $\delta^{\rm (pot)}_{2}$.
Recently, the microscopic CC calculation with a complex $G$-matrix interaction has been successfully applied to heavy-ion scattering~\cite{TAK10, WWQ15, WWQ17}.
The power of the complex $G$-matrix interaction is also shown not only in reproducing the experimental data but also in predicting interesting nuclear reaction phenomena~\cite{FUR10, FUR13-1, FUR18, FUR21-2}.

In this study, we aim to elucidate the relationship between $\delta_{2}$ and $\delta^{\rm (pot)}_{2}$
using the DD and DP models in a wide range of incident energies and target nuclei.
The earlier studies investigated this relationship
mainly for the lower-energy region ($E/A <$ 100 MeV),
and showed that the use of the DP model significantly underestimates the nuclear deformation length~\cite{BEE96, KHO00}.
Extending these analyses to the high-energy region is challenging
due to the lack of experimental data on high-energy heavy-ion scattering.
As mentioned, recent development of the folding model approaches allows us to make a reliable prediction of high-energy heavy-ion scattering.
Therefore, it is worthwhile to proceed with a theoretical analysis in the high-energy region. 
It should be noted that we can discuss deformation effects more straightforwardly
as the reaction mechanism becomes simpler at higher incident energies.
In the present study, we consider the $^{12}$C inelastic scattering
by $^{12}$C, $^{16}$O, $^{40}$Ca, and $^{208}$Pb targets at $E/A =$ 50--400 MeV.

This paper is organized as follows.
In Sec.~\ref{sec:formalism}, we explain how to relate $\delta_{2}$ with
$\delta^{\rm (pot)}_{2}$ using the DD and DP models.
This section is further divided into three subsections.
In Sec.~\ref{sec:MP}, we present the theoretical framework to obtain
the microscopic potential, which is used in both the DD and DP models.
The DD and DP models are detailed in Secs.~\ref{sec:DD} and \ref{sec:DP}, respectively.
In Sec.~\ref{sec:results}, we first show the validity of the present models
for the elastic scattering.
Next, we calculate the angular-integrated inelastic scattering cross sections using the DD model.
The cross sections are used as reference calculations to extract $\delta^{\rm (pot)}_{2}$ in the DP model.
Then, $\delta_{2}$ and $\delta^{\rm (pot)}_{2}$ 
are compared systematically.
Lastly, the conclusion is given in Sec.~\ref{sec:conclusion}.

\section{Formalism}
\label{sec:formalism}

We calculate the inelastic scattering cross sections for the $2_1^+$ state of $^{12}$C, denoted as $\sigma (2_1^+)$, using the CC formalism.
In the CC calculation, both diagonal and coupling potentials are required. 
They are obtained with two models: the deformed density (DD) model and the deformed potential (DP) model.
In the DD model, we first assume a deformed density characterized by $\delta_{2}$.
Then, we can construct the diagonal and coupling potentials microscopically through the folding procedure.
Once these potentials are determined, the $\sigma (2_1^+)$ can be calculated in the standard CC framework.
The result of the DD model is used as a reference calculation in this paper.
On the other hand, in the DP model, we derive the coupling potential by assuming a deformed potential
characterized by $\delta^{\rm (pot)}_{2}$. The value of $\delta^{\rm (pot)}_{2}$ is determined
so as to reproduce the $\sigma (2_1^+)$ calculated with the DD model.
Finally, we systematically compare $\delta^{\rm (pot)}_{2}$ with $\delta_{2}$ in high-energy heavy-ion scattering and elucidate the relationship between them.

\subsection{Microscopic potentials}\label{sec:MP}
We briefly summarize the construction of the microscopic potential used in this paper.
The detailed formulation is described in Refs.~\cite{ITO01, KAT02, FUR13-1, FUR13-2}.

We consider the scattering of a deformed projectile (P) and a spherical target (T).
The diagonal and coupling potentials between P and T are obtained
by folding the effective nucleon-nucleon interaction $v_{NN}$ with the projectile and target densities:
\begin{align}
U^{(\lambda)}_{I' I}(R) 
&= \int \rho^{(\lambda)}_{I' I} (r_1) \rho_\mathrm{T}(r_2)v_{NN}(s,\rho) \nonumber\\
&\hspace{5mm}\times\left[\mathscr{Y}_\lambda(\hat{\bm{r}}_1) \otimes \mathscr{Y}_\lambda(\hat{\bm{R}}) \right]_{00} d\bm{r}_1 d\bm{r}_2 d\hat{\bm{R}} \label{eq:dfm},
\end{align}
where $\bm{R}$ is the coordinate between P and T, $\bm{r}_1$ ($\bm{r}_2$) is the coordinate of the interacting nucleon from the center of mass of P (T), and $\bm{s}=\bm{R}-\bm{r}_1+\bm{r}_2$.
The subscripts $I$ and $I'$ are the initial and final spins of P, respectively, $\lambda$ denotes the multipolarity, and $\mathscr{Y}_{LM}(\hat{\bm{r}})=i^L Y_{LM}(\hat{\bm{r}})$.
The density-dependent part $\rho$ in $v_{NN}$ is taken based on
the frozen density approximation~\cite{FUR09, FUR10}.
The validity of this approximation is discussed in Refs.~\cite{KHO07, FUR09, FUR16}.
We define the transition density as
\begin{equation}
    \rho_{I'm',Im}(\bm{r}) = \sqrt{4\pi}\sum_{\lambda \mu}{(Im\lambda \mu |I'm')
\ \rho^{(\lambda )}_{I' I}(r) \mathscr{Y}_{\lambda \mu}^*(\hat{\bm{r}})}, \label{eq:trden}
\end{equation}
where $\rho^{(\lambda)}_{I' I}$ is the radial part of the transition density with rank $\lambda$,
$(Im\lambda \mu |I'm')$ is the Clebsch-Gordan coefficient,
and $m$, $m'$, and $\mu$ denote the $z$ components of $I$, $I'$, and $\lambda$, respectively.
In Eq.~\eqref{eq:dfm}, T is assumed to be inert, that is, $\rho_\mathrm{T}(r_2)$ is the ground state density $\rho^{(\lambda =0)}_{00}(r_2)$ in the definition~\eqref{eq:trden}.
Equation~(\ref{eq:dfm}) is the so-called direct part,
and the exchange part is similarly obtained in the folding procedure as prescribed in Refs.~\cite{KHO00, FUR13-1, FUR13-2}.
The Coulomb potential is also constructed in the folding model, where the Coulomb nucleon-nucleon interaction is folded with the proton densities.

In the actual calculation, we adopt the complex $G$-matrix interaction MPa~\cite{YAM14}
for the effective nucleon-nucleon interaction in Eq.~(\ref{eq:dfm}).
The MPa interaction is derived from the realistic nucleon-nucleon interaction~\cite{RIJ061, RIJ062} in the $G$-matrix calculation.
The MPa interaction satisfies the saturation property in the infinite nuclear matter by applying the three-nucleon force.
Since the complex $G$ matrix is constructed for infinite nuclear matter, the strength of its imaginary part is often adjusted for the finite nucleus because their level densities are quite different.
Therefore, we apply the incident-energy-dependent renormalization factor,
$N_W = 0.5 + (E~[{\rm MeV}]/A)/(1000~[{\rm MeV}])$~\cite{FUR19}, to the imaginary part.
We note that no additional parameter is introduced.
Consequently, once the transition densities are determined, the elastic and inelastic 
scattering cross sections can be uniquely calculated from the double-folding potentials.
Relativistic kinematics is used.

\subsection{Deformed density model}\label{sec:DD}
For the reference calculations of inelastic scattering cross sections, we employ the deformed density (DD) model.
To make the discussion clearer, we consider the deformed Fermi-type (DF) density in the first-order approximation~\cite{KHO00}:
\begin{align}
\rho^{(\lambda=0)}_{\rm in}(r') &= \sqrt{4\pi} \rho_{\rm DF}(r'), \\ \rho^{(\lambda=2)}_{\rm in}(r') &= - \delta_{2} \frac{d\rho_{\rm DF}(r')}{dr'} \label{eq:rho2}
\end{align}
with
\begin{equation}
\rho_{\rm DF}(r') = \frac{\rho_0}{1+\exp{\left( \frac{r'-R_0}{a} \right)}},
\end{equation}
where $\rho_0$ is the normalization constant, $R_0$ and $a$ are the radius and the diffuseness parameters, respectively.
According to Ref.~\cite{KAM81}, the intrinsic density $\rho^{(\lambda)}_{\rm in}$ can be transformed into $\rho^{(\lambda )}_{I' I}$ in Eq.~\eqref{eq:dfm} as
\begin{equation}
\rho^{(\lambda )}_{I' I}(r) = \frac{i^\lambda}{\sqrt{4\pi}} \rho^{(\lambda)}_{\rm in} (r) (I'0\lambda 0 |I 0).
\end{equation}
Using the transition densities defined above,
we calculate the inelastic scattering cross section,
which is used as a reference for the following analysis.

\subsection{Deformed potential model}\label{sec:DP}
Another way of calculating inelastic scattering cross sections is based on the deformed potential (DP) model, which has been conventionally used in the analysis of experiments.
The deformed potential $U_{\rm DP}(R, \theta')$ can be expanded as 
\begin{equation}
U_{\rm DP}(R, \theta') = \sum_{\lambda} {U_{\rm DP}^{(\lambda)}(R) Y_{\lambda 0}(\theta')}, \label{eq:dp}
\end{equation}
where $U_{\rm DP}^{(\lambda)}$ is the radial part of the deformed potential,
$\theta'$ is the direction of the target nucleus in the intrinsic frame. For simplicity, we consider the first-order approximation of $U_{\rm DP}$ as
\begin{align}
U_{\rm DP}^{(\lambda=0)}(R) &= \sqrt{4\pi} U(R) \label{eq:DP_lam0},
\\ U_{\rm DP}^{(\lambda=2)}(R) &= - \delta_2^{\rm (pot)} \frac{d U(R)}{dR}, \label{eq:DP_lam2}
\end{align}
where $\delta^{\rm (pot)}_{2}$ is the potential deformation length,
and $U$ is the optical potential, for which the Woods-Saxon form is often taken.
In the present analysis, we apply the microscopic optical potential obtained in Eq.~\eqref{eq:dfm} to $U$, i.e.,
\begin{equation}
U(R)=U^{(\lambda=0)}_{00}(R).\label{eq:U00}
\end{equation}
This procedure ensures a fair comparison between the DD and DP models
by maintaining the common potential for the entrance channels in both models.
For simplicity, the values of $\delta^{\rm (pot)}_{2}$ for the real and imaginary parts are taken to be identical.
Furthermore, the Coulomb coupling potential is taken to be the same as in the DD model.

\section{Results}
\label{sec:results}

We compare $\delta_{2}$ with $\delta^{\rm (pot)}_{2}$ for the $^{12}$C inelastic scattering by the $^{12}$C, $^{16}$O, $^{40}$Ca, and $^{208}$Pb targets at $E/A =$ 50--400 MeV.
We search for the optimal $\delta^{\rm (pot)}_{2}$ so as to reproduce the inelastic scattering cross sections for the 2$^+_1$ state of $^{12}$C, $\sigma (2_1^+)$, obtained in the DD model with $\delta_{2}$. 
For the deformed nucleus $^{12}$C, $R_0=2.1545$ fm, $a=0.425$ fm, and $|\delta_{2}|=1.564$ fm are employed as used in Ref.~\cite{KHO00}, which reproduce the experimental $B(E2)$ value~\cite{RAM87}.
We take $\delta_2<0$ as $^{12}$C has an oblate shape.
The ground-state densities of $^{16}$O, $^{40}$Ca, and $^{208}$Pb are obtained from the Hartree-Fock calculations available in Ref.~\cite{NEG70}. 
The nuclear excitation of doubly-magic nuclei is neglected as its effect is expected to be small.
For the $^{12}$C + $^{12}$C system, the symmetrization of 
identical particles is made but only a single (target or projectile) excitation is considered.

\begin{figure}[ht]
\centering
\includegraphics[width=0.9\linewidth]{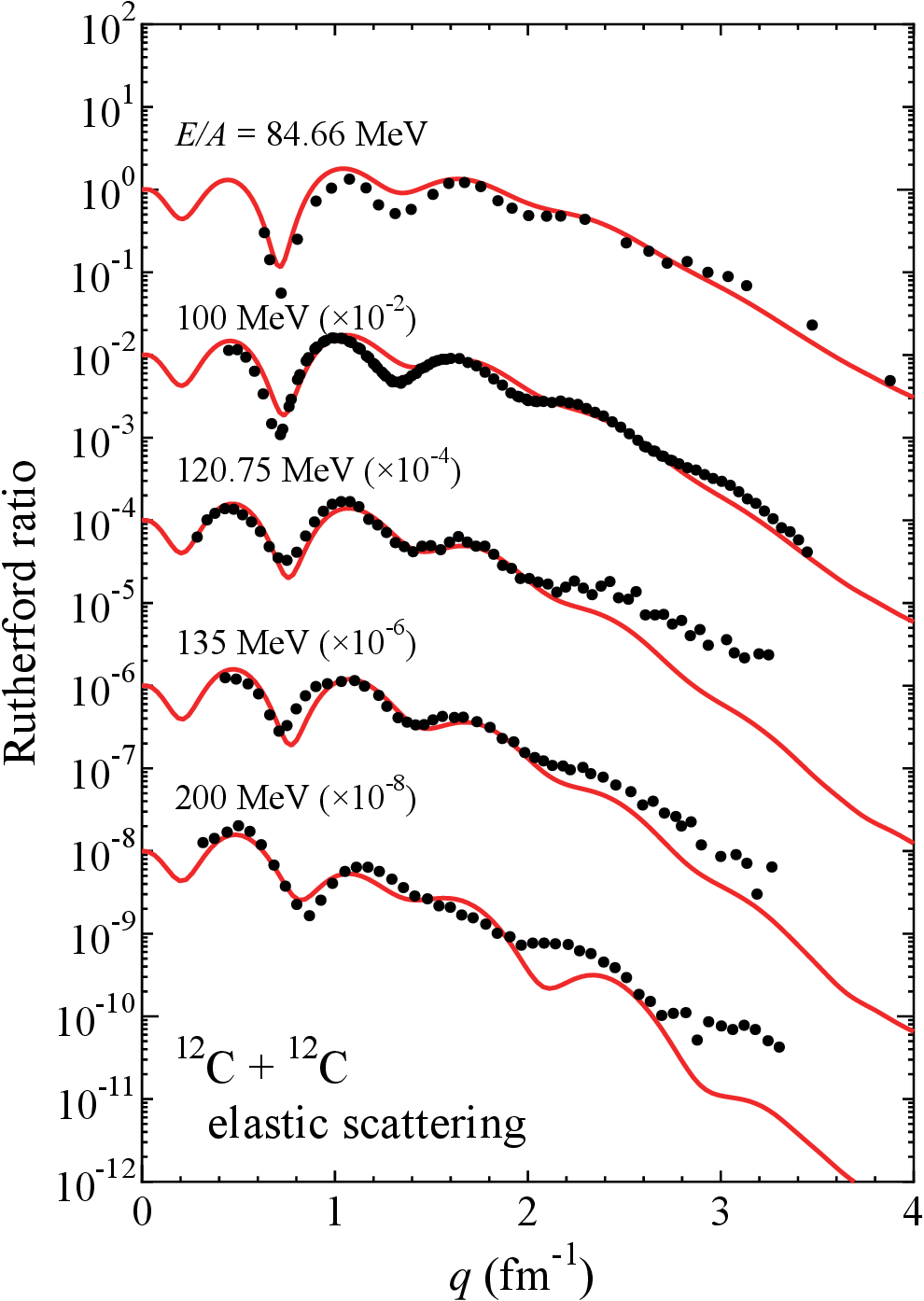}
\caption{Angular distributions of elastic scattering for the $^{12}$C + $^{12}$C system at $E/A$ = 85--200 MeV.
The experimental data are taken from Refs.~\cite{WWQ15, WWQ17, BUE81, ICH94, HOS87}.
}
\label{fig:el-cc}
\end{figure}

\begin{figure}[ht]
\centering
\includegraphics[width=0.9\linewidth]{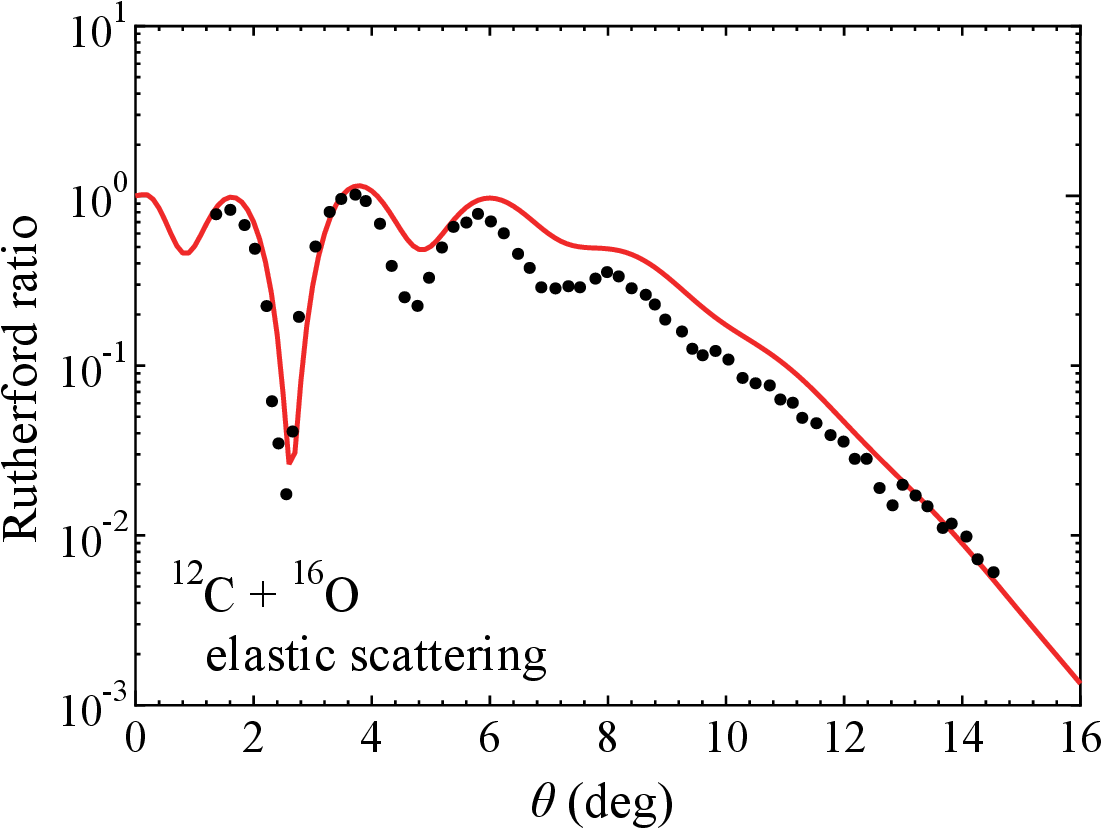}
\caption{Angular distributions of elastic scattering for the $^{12}$C + $^{16}$O system at $E/A$ = 94 MeV.
The experimental data are taken from Ref.~\cite{ROU88}.
}
\label{fig:el-oc}
\end{figure}
First, we show the validity of microscopic potentials
 by comparing the DD model with available experimental data for elastic scattering cross sections.
Figure~\ref{fig:el-cc} illustrates the angular distributions of the elastic scattering cross sections for
the $^{12}$C + $^{12}$C system at $E/A$ = 85--200 MeV.
The solid curves represent the microscopic CC calculations
and reproduce the experimental data.~\footnote{The MPa $G$-matrix interaction used in this paper is slightly modified from that in Ref.~\cite{YAM14}
by some improvements for numerical computations and minor error corrections in the $G$-matrix code. Therefore, the present results are different from those in Ref.~\cite{FUR16}, especially at $E/A$ = 200 MeV.}
Similarly, in Fig.~\ref{fig:el-oc}, the elastic scattering cross sections for $^{12}$C + $^{16}$O at $E/A$ = 94 MeV 
are also reproduced well, especially in the forward angles.
We see the reliability of the diagonal potential $U^{(\lambda=0)}_{00}$ used in both the DD and DP models and proceed with the present analysis.

\begin{figure}[ht]
\centering
\includegraphics[width=0.9\linewidth]{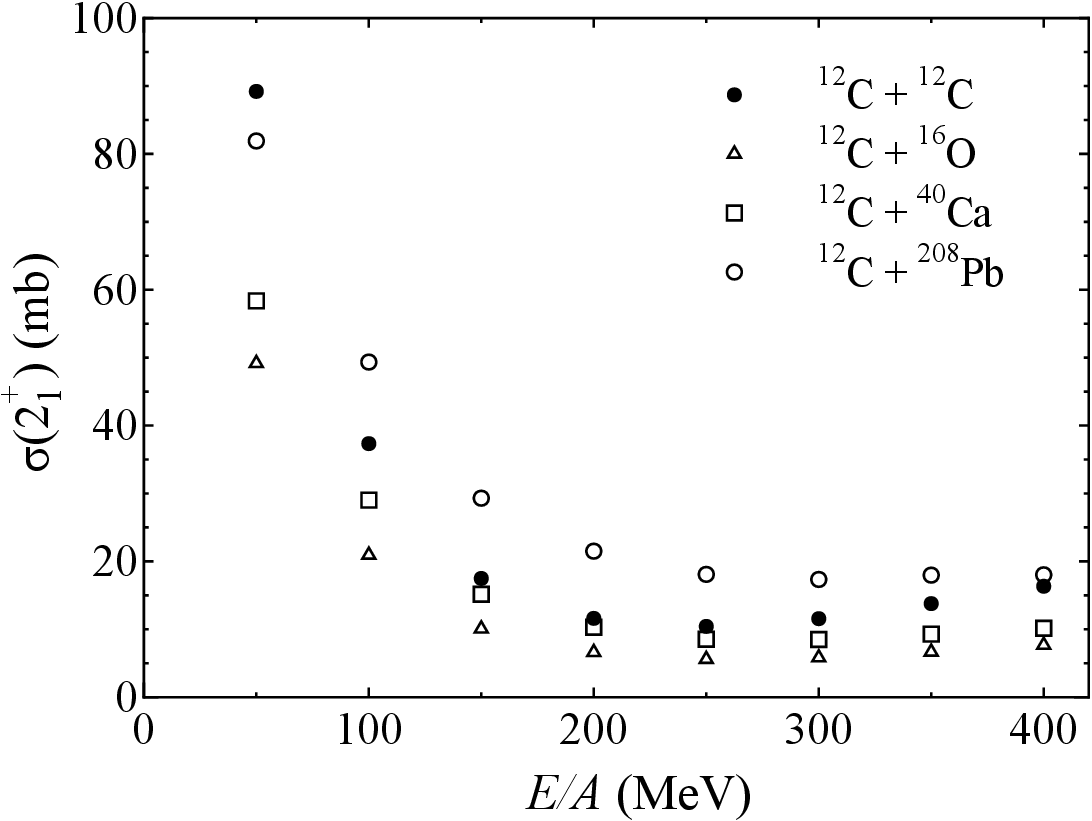}
\caption{Angular-integrated inelastic scattering cross sections for the $2^+_1$ state of the $^{12}$C nucleus [$\sigma (2_1^+)$]
 by the $^{12}$C (filled circles), $^{16}$O (open triangles),
$^{40}$Ca (open squares), and $^{208}$Pb (open circles) targets
 at incident energies $E/A = 50$--400 MeV.
 These results are obtained from the DD model and used as reference calculations for determining $\delta^{\rm (pot)}_2$ in the DP model.
}
\label{fig:TCS_all}
\end{figure}

Next, we calculate the angular-integrated inelastic scattering cross sections $\sigma (2_1^+)$
as reference calculations using the DD model.
Figure~\ref{fig:TCS_all} shows the $\sigma (2_1^+)$ for $^{12}$C scattering
by the $^{12}$C, $^{16}$O, $^{40}$Ca, and $^{208}$Pb targets at $E/A$ = 50--400 MeV.
The filled circles, open triangles, open squares, and open circles correspond
to the reactions by $^{12}$C, $^{16}$O, $^{40}$Ca, and $^{208}$Pb targets, respectively.
The $\sigma (2_1^+)$ rapidly decreases as the incident energy increases up to $E/A \lesssim$ 200 MeV.
We find that, in the low-energy region ($E/A \lesssim$ 100 MeV), the real part of the coupling potentials plays a decisive role in determining the $\sigma (2_1^+)$
because the imaginary part is relatively small.
The strength of the real part becomes weaker as the energy increases;
it is noteworthy that the real part of the diagonal potential shows the repulsive nature at $E/A \sim$ 200 MeV.
Beyond this energy ($E/A \gtrsim$ 250 MeV), the $\sigma (2_1^+)$ exhibits a weaker dependence
on the incident energy. In the high-energy region, the imaginary part of the coupling potentials plays a major role in the $\sigma (2_1^+)$ values.
We find that the contribution of the imaginary part to $\sigma (2_1^+)$ is almost constant in the energy range of our analysis ($50 \le E/A \le 400$ MeV).
It should be noted that the $\sigma (2_1^+)$ for $^{12}$C + $^{12}$C scattering is relatively large although the target mass is the smallest.
This is because the symmetrization procedure involving the single excitation is taken into account for this system.
These theoretical results are used as reference calculations
for determining $\delta^{\rm (pot)}_2$ in the DP model.

For $^{12}$C + $^{12}$C inelastic scattering at $E/A \sim 100$ MeV,
several experimental data are available.
At $E/A = 121.1$ MeV, our calculation of $\sigma_\mathrm{theo} (2_1^+)=27$ mb underestimates the observed data of $\sigma_\mathrm{exp} (2_1^+)=43\pm3$ mb~\cite{TAKECHI09,EXFOR_TAKECHI09}.
Conversely, for the angular distribution $d\sigma (2_1^+)/d\Omega$ at $E/A = 100$ MeV, our result tends to overestimate the experimental data as was also shown in Refs.~\cite{WWQ15, WWQ17}, which employed a similar reaction model.
This discrepancy highlights the need for further investigations.
Measurements of inelastic cross sections for heavy-ion scattering could provide
crucial insights for the quantitative refinement of microscopic potentials.

\begin{figure}[ht]
\centering
\includegraphics[width=0.9\linewidth]{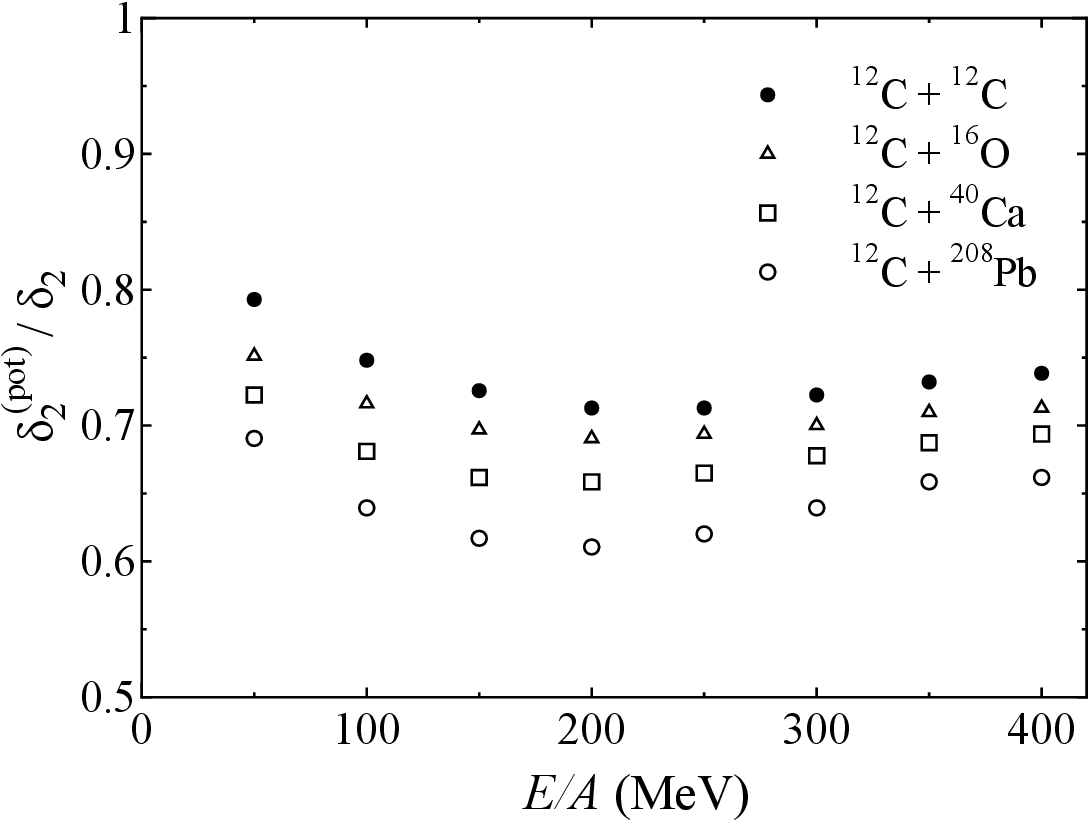}
\caption{Deformation length of the nuclear potential $\delta^{\rm (pot)}_2$ derived from $^{12}$C inelastic scattering cross sections at $E/A =$ 50--400 MeV, divided by $\delta_2=-1.564$ fm.
The filled circles, open triangles, open squares, and open circles
represent the results for the scattering by $^{12}$C, $^{16}$O, $^{40}$Ca, and $^{208}$Pb targets, respectively.
}
\label{fig:dl-12c}
\end{figure}

Figure~\ref{fig:dl-12c} illustrates the energy dependence of $\delta^{\rm (pot)}_2$
derived from the $\sigma (2_1^+)$ calculated with the DD model.
Note that the values of $\delta^{\rm (pot)}_2$ are divided by $\delta_2$.
The filled circles, open triangles, open squares, and open circles
represent the results for the scattering of $^{12}$C
by $^{12}$C, $^{16}$O, $^{40}$Ca, and $^{208}$Pb targets, respectively.
Our primary finding is that $\delta^{\rm (pot)}_2$ 
gets overall underestimation, which is approximately 20--40 \% smaller than $\delta_2$,
and shows strong incident energy and target dependence.
The $\delta^{\rm (pot)}_2$ values become smaller as the target mass increases.
We confirmed that this behavior is kept even when the folding potential
for the elastic channel [$U^{(\lambda=0)}_{00}$ in Eq.~\eqref{eq:U00}] is replaced
 with a phenomenological Woods-Saxon potential that is determined to reproduce the elastic scattering cross section calculated with the DD model.
This significant deviation casts doubt on
the determination of $\delta_2$ with $\delta^{\rm (pot)}_2$ from the high-energy heavy-ion scattering.
A systematic underestimation of the quadrupole deformation length is expected
in studies based on the DP model that assumes 
$\delta_2=\delta^{\rm (pot)}_2$.

\begin{figure}[ht]
\centering
\includegraphics[width=0.9\linewidth]{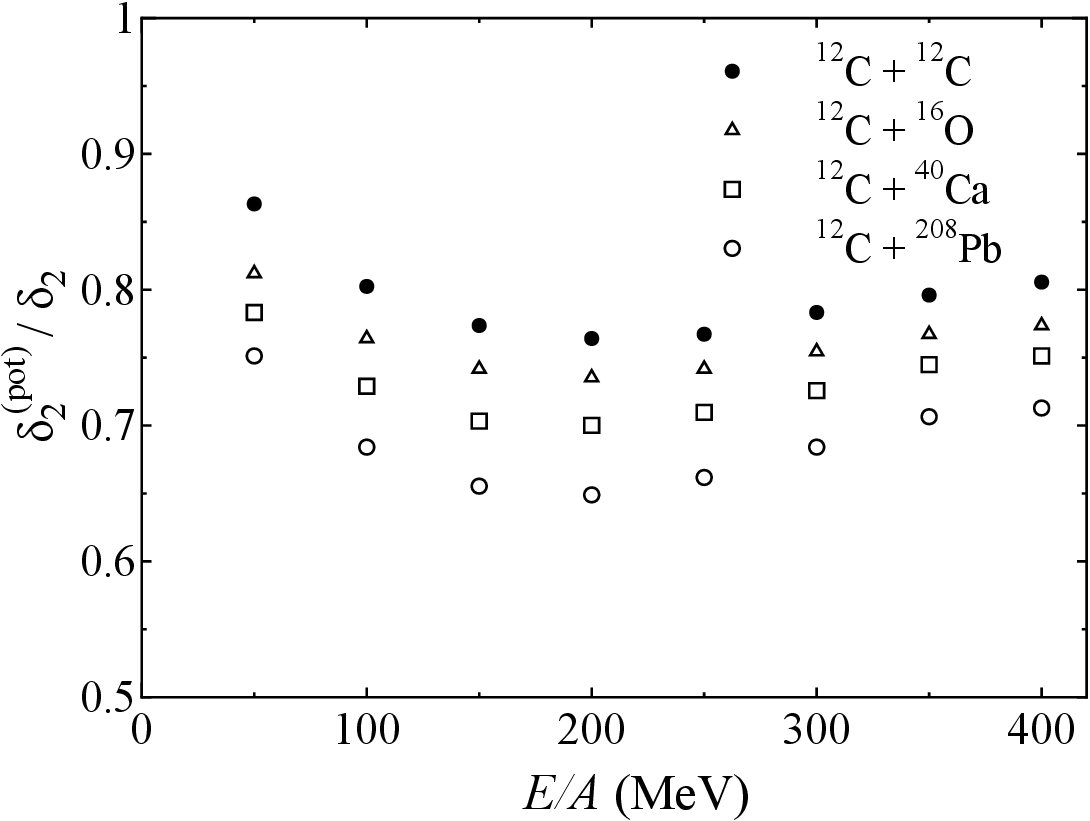}
\caption{Same as Fig.~\ref{fig:dl-12c} but $\delta^{\rm (pot)}_2$ is
extracted from the DP2 model.
}
\label{fig:dl-12c_DP2}
\end{figure}

Lastly, we investigate the higher-order effect of $\delta^{\rm (pot)}_2$ on
the form factor $U_{\rm DP}^{(\lambda=2)}(R)$ beyond the derivative form
given in Eq.~\eqref{eq:DP_lam2}, because the extracted $\delta^{\rm (pot)}_2$ is relatively large
($|\delta^{\rm (pot)}_2| \sim 1.1$ fm).
In considering the folding potential, we modify the optical potential as
\begin{equation}
U_{\rm DP2}(R, \theta') =U\left(R-\delta^{\rm (pot)}_2 Y_{20}(\theta')\right),\label{eq:Udp2}
\end{equation}
where $U$ is the arbitrary optical potential [Eq.~\eqref{eq:U00}], and
the subscript ``DP2" denotes the deformed potential 2 model, which is distinguished from the DP (derivative) model.
This method is commonly used, for example, in FRESCO~\cite{THO88}.
Following Eq.~(16) of Ref.~\cite{TAM65}, we further define the form factor in the DP2 model
\begin{equation}
U_{\rm DP2}^{(\lambda=2)}(R) = 4\pi \int_0^1
U_{\rm DP2}(R, \theta')Y_{20}(\theta')\,d(\cos{\theta'}). \label{eq:DP2_lam2}
\end{equation}
In the present analysis, we assume the monopole part as $U_{\rm DP2}^{(\lambda=0)}(R) = \sqrt{4\pi} U(R)$ to maintain the consistency of the DD and DP models.
Note that Eq.~\eqref{eq:DP2_lam2} reduces to Eq.~\eqref{eq:DP_lam2} when $|\delta^{\rm (pot)}_2|$ is small.
Figure~\ref{fig:dl-12c_DP2} shows $\delta^{\rm (pot)}_2$ extracted using the DP2 model,
where the values of $\delta^{\rm (pot)}_2$ are divided by $\delta_2$ as in Fig.~\ref{fig:dl-12c}.
The overall trend is the same as Fig.~\ref{fig:dl-12c} but the ratios are increased by 6--9 \%
from the DP model.
This discrepancy arises only from $U_{\rm DP2}^{(\lambda=2)}$, whose peak position
slightly shifts inward for larger $|\delta^{\rm (pot)}_2|$ compared to $U_{\rm DP}^{(\lambda=2)}$.
This behavior results in smaller $\sigma (2_1^+)$ for the same value
of $|\delta^{\rm (pot)}_2|$, leading to the extraction of larger $|\delta^{\rm (pot)}_2|$ in the DP2 model.
However, the extracted $\delta^{\rm (pot)}_2$ remains 15--35 \% smaller than $\delta_2$,
indicating significant underestimation even in the DP2 model.

\section{Conclusion}
\label{sec:conclusion}

We have investigated the relation between the quadrupole deformation lengths of the nuclear density and potential ($\delta_{2}$ and $\delta^{\rm (pot)}_{2}$) for the $^{12}$C inelastic scattering by the $^{12}$C, $^{16}$O, $^{40}$Ca and $^{208}$Pb targets at $E/A$ = 50--400 MeV.
For this analysis, we employ two models: the deformed density (DD) model and the deformed potential (DP) model.
In the DD model, the coupling potential is microscopically
constructed from the transition density based on the deformed density characterized by $\delta_{2}$.
In the DP model, the coupling potential is derived 
based on the deformed potential characterized by $\delta^{\rm (pot)}_{2}$, 
which is determined to reproduce the inelastic scattering 
cross section calculated with the DD model.
We find that $\delta^{\rm (pot)}_{2}$ shows 
overall underestimation of $\delta_{2}$ by 20--40 \%,
having strong incident energy and target dependence.
Further analysis using the DP2 model, which considers higher-order deformation effects beyond
the DP (derivative) model, reveals that $\delta^{\rm (pot)}_{2}$ is still about 15--35 \% smaller than $\delta_{2}$.
These results clearly indicate that the assumption $\delta_2=\delta^{\rm (pot)}_2$ is too naive
for the determination of the nuclear deformation 
using the high-energy heavy-ion scattering in the DP model.

\vspace{2mm}
\acknowledgments
This work was supported by Research Network Support Program, National Institute of Technology (KOSEN) and Japan Society for the Promotion of Science (JSPS) KAKENHI Grants  No. JP18K03635, No. JP20K03943, No. JP21K03543, No. JP22K03610, No. JP22K14043, No. JP22H01214, and No. JP23K22485.
One of the authors, S.W., thanks A. M. Moro and the faculty and staff
at Universidad de Sevilla for their hospitality during
his sabbatical stay, which enabled the completion of this work.

\bibliography{WF-deformation}

\end{document}